\documentstyle[12pt]{article}
\renewcommand{\section}[1]{\addtocounter{section}{1}
       \vspace{5mm} \par \noindent
        {\bf \thesection . #1}\setcounter{subsection}{0}
        \par \vspace{2mm} } 
\renewcommand{\subsection}[1]{\addtocounter{subsection}{1}
         \vspace{2.5mm}\par\noindent {\em \thesubsection . #1}\par
          \vspace{0.5mm} }
\renewcommand{\thebibliography}[1]{
          { \vspace{5mm}\par\noindent
          {\bf References}\par\vspace{4mm} } \list{[\arabic{enumi}]}    
          {\settowidth\labelwidth{[#1]}\leftmargin
          \labelwidth \advance\leftmargin\labelsep\addtolength{\topsep}{-4em}
          \usecounter{enumi}}  }

\newcommand{\hoch}[1]{$\, ^{#1}$}
\newcommand{\eqn}[1]{eq.(\ref{#1})}

\def\ft#1#2{{\textstyle{{\scriptstyle #1}\over {\scriptstyle #2}}}}

\def\del{\partial}
\def\a{\alpha}

\def\g{\gamma}
\def\t{\theta}
\def\ad{\dot\alpha}

\def\1d{\dot 1}
\def\2d{\dot 2}    

\def\ni{\noindent}

\def\bm{\bibitem}
\def\be{\begin{equation}} 
\def\ee{\end{equation}} 
\def\ba{\begin{eqnarray}} 
\def\ea{\end{eqnarray}}

\begin{document}
\begin{flushright}
\hfill{CTP TAMU-6/96}\\
\hfill{hep-th/9602099}\\
\end{flushright}

\vspace{30pt} 
\begin{center}

{\Large\bf Is There a Stringy Description of Self-Dual Supergravity in
           2+2 Dimensions? }\\

\vspace{40pt}

{\large E. Sezgin\hoch{\dagger}} \\

\vspace{15pt}

{\it Center for Theoretical Physics \\
Texas A\&M University, College Station, TX 77843}

\vspace{50pt}         

\centerline{ABSTRACT}
\end{center}

The well known N=2 string theory describes self-dual gravity, as was
shown by Ooguri and Vafa sometime ago. In search of a variant of this
theory which would describe self-dual supergravity in 2+2 dimensions, we
have constructed two new N=2 strings theories in which the target space
is a superspace. Both theories contain massless scalar and spinor fields
in their spectrum, and one of them has spacetime supersymmetry. However,
we find that the interactions of these fields do not correspond to those
of self-dual supergravity. In our construction, we have used the basic 
(2,2) superspace variables, and considered quadratic constraints in
these variables. A more general construction may be needed for a stringy
description of self-dual supergravity.

{
\vfill\leftline{}\vfill\vskip 10pt\footnoterule
{\footnotesize\hoch{\ast}
Talk presented at the Workshop on {\it Strings, Gravity and
Related Topics}, Trieste, Italy, \hfill\break\phantom\quad\quad\quad
June 29-30, 1995.}
\vskip -12pt
}
\vskip 10pt
{\footnotesize\hoch{\dagger}
Research supported in part by NSF Grant PHY-9411543}

\pagebreak
\setcounter{page}{1}

\section{Introduction}

Various supergravity theories are known to arise as low energy effective
field theory limits of an underlying superstring or super $p$-brane
theory. For example, all supergravity theories in $D=10$ and $D=11$
are associated with certain superstring or super $p$-brane theories.
Supergravity theories can also serve as worldvolume field theories for
a suitable super $p$-brane theory, the most celebrated example of this
being the spinning string theory.
 
Of course, not all supergravity theories have been associated so far
with superstrings or super $p$-branes. An outstanding example is the
self-dual supergravity in $2+2$ dimensions \cite{bs,kgn,ws1}. There are a
number of reasons why this is a rather important example. For one thing,
the dimensional reduction to $1+1$ dimensions can give rise to a large
class integrable models. Secondly, it can teach us a great deal about
quantum gravity. Furthermore, and perhaps more interestingly, a suitable
version of self-dual supergravity in $2+2$ dimensions may in principle
serve as the worldvolume theory of an extended object  propagating in $10+2$
dimensions, as has been suggested recently by Vafa \cite{vafa}.  Further
tantalizing hints at the relevance of a worldvolume theory in $2+2$ dimensions
have been put forward recently \cite{km}.

Since the well known $N=2$ string theory has the critical dimension of
four, it is natural to examine this theory, or its variants, in search of
a stringy description of self-dual supergravity. It turns out that this
theory actually describes self-dual gravity in 2+2 dimensions, as was
shown by Ooguri and Vafa \cite{ov} sometime ago.  Interestingly enough,
and contrary to what one would naively expect, the fermionic partner of
the graviton does not arise in the spectrum, and therefore self-dual
supergravity does not emerge \cite{ov}. This intriguing result led us to
look for a variant of the $N=2$ string theory where spacetime
supersymmetry is kept manifest from the outset, thereby providing a
natural framework for finding a stringy description of self-dual
supergravity. We have constructed two such variants \cite{us1,us2}, in
which (a) we use the basic variables of the $2+2$ superspace, and (b) we
consider constraints that are quadratic in these variables. 
Surprizingly enough, we find that neither one of the two models describe
the self-dual supergravity, suggesting that we probably need to
introduce extra world-sheet variables and/or consider higher order
constraints. Nonetheless, we believe that our results may be of interest
in their own right, and with that in mind, we shall briefly describe
them in this note.

Both of the models mentioned above can be constructed by making use of
bilinear combinations of the bosonic coordinates $X^{\a\ad}$, fermionic
coordinates $\t^\a$, and their conjugate momenta $p_\a$, to built the
currents of the underlying worldsheet algebras. The indices $\a$ and
$\ad$ label the two dimensional spinor representations of $SL(2)_R\times
SL(2)_L \approx SO(2,2)$. In terms of these variables, it is useful to
recall the currents of the small $N=4$ superconformal algebra, namely
\begin{eqnarray}
T&=& -\ft12 \del
X^{\alpha\dot\alpha}\del X_{\alpha\dot\alpha}
     -p_\alpha \del \theta^\alpha \ , \nonumber\\
G^{\dot \alpha} &=& p_{\alpha}\del X^{\alpha\dot \alpha}\ , \quad
{\widetilde G}^{\dot \alpha} = \theta_{\alpha}\del X^{\alpha\dot \alpha}
 \ , \label{n4alg}\\
 J_0 &=& p_{\alpha}\theta^{\alpha}\ , \quad J_+= p_\alpha p^\alpha\ ,
\quad J_-=\theta_\alpha\theta^\alpha\ . \nonumber
\end{eqnarray}

\ni This is the twisted version of the usual realization, since here the
$(p,\t)$ system has dimension $(1,0)$. An $N=2$ truncation of this
algebra is given by \cite{us1}
\begin{eqnarray}
T&=& -\ft12 \del X^{\alpha\dot\alpha}\del X_{\alpha\dot\alpha}
     -p_\alpha \del \theta^\alpha \ , \qquad
 G^{\dot \alpha} = p_{\alpha}\del X^{\alpha\dot\alpha} \ ,
                    \qquad J = p_\alpha p^\alpha \ . \label{tn4alg}
\end{eqnarray}

\ni Naively, this system appears to be non-critical. However,
the currents are reducible, and a proper quantization requires the
identification of the irreducible subsets. Assuming that
\begin{enumerate}
\item[(a)]
 the worldsheet field content is $(p_\a, \t, X^{\a\ad})$,
\item[(b)]
 the constraints are quadratic in worldsheet fields, 
\item[(c)]
 the constraints are irreducible,
\end{enumerate}

\ni  we have found that, there exists three possible $N=2$ string
theories. One of them is the old model shown by Ooguri and Vafa
\cite{ov} to descibe pure self-dual gravity. We will refer to this model
as the ``$n=0$ model''. The other two models were studied in refs.
\cite{us1,us2}. One of them, which we will refer to as the ``$n=1$
model'', has {\it spacetime} $N=1$ supersymmetry, and the other one,
which we will refer to as the ``new $n=0$ model'', has no spacetime
supersymmetry. In what follows, we shall give a very brief description
of these models.

\section{The $N=2$ String Models}

\subsection{The $n=0$ Model}

   This is the usual $N=2$ string which has worldsheet $N=2$
supersymmetry, but lacks spacetime supersymmetry. The underlying $N=2$
superconformal algebra, in the twisted basis described above, is given by
\begin{eqnarray}
T&=& -\ft12 \del X^{\alpha\dot\alpha}\del X_{\alpha\dot\alpha}
     -p_\alpha \del \theta^\alpha \ , \qquad J=p_{\alpha}\theta^{\alpha}\ ,
     \nonumber\\
     G^{\dot 1} &=& \theta_{\alpha}\del X^{\alpha\dot 1}\ , \qquad
     G^{\dot 2} = p_{\alpha}\del X^{\alpha\dot 2} \ .  \label{n2alg}
\end{eqnarray}

The striking feature of this model is that the only continuous degree of
freedom it describes is that of the self-dual graviton \cite{ov}.
This model has been studied extensively in the literature. See, for
example, refs.~\cite{lp,l}, where a BRST analysis of the spectrum is
given, and various twists and GSO projections leading to massless
bosonic and fermionic vertex operators are considered. We now turn our
attention to the remaining two models, which we have constructed in
\cite{us1,us2}.

\subsection{The n=1 Model}

This model can covariantly be described by the set of currents given in
\eqn{tn4alg}
\footnote{ In \cite{ws2}, Siegel proposed to build a string theory
implementing the set of constraints given by $\Big\{\del
X^{\alpha\dot\alpha}\del X_{\alpha\dot\alpha},\,
p_\alpha\,\del\theta^\alpha, \,p_\alpha p^\alpha,\, \del\theta_\alpha\,
\del\theta^\alpha,\, p_\alpha\,\del X^{\alpha\dot\alpha},\,
\del\theta_\alpha\, \del X^{\alpha\dot\alpha}\Big\}$.  However, we have
checked that the algebra of these constraints does not close \cite{us1}.
Actually, this non-closure occurs even at the classical level of Poisson
brackets, or single OPE contractions \cite{us1}.}.
Notice that all currents have spin two, and that the system
is critical. Nonetheless, this set of constraints is reducible. All the
relations among the constraints can be described in a concise form by
introducing a pair of spin-0 fermionic coordinates $\zeta^{\dot\a}$ on
the worldsheet.  We can then define
\begin{equation}
{\cal P}^\a= p^\a +\zeta_{\dot\a}\, \del X^{\a\dot\a} + \zeta_{\dot\a}\,
 \zeta^{\dot\a}\, \del\theta^\alpha\ ,\label{psuper}
\end{equation}

\ni in terms of which the currents may be written as ${\cal T}={\cal P}_\a\, 
{\cal P}^\a$, where
\begin{equation}
{\cal T}= J + \zeta_{\dot\a}\, G^{\dot\a} +
\zeta_{\dot\a}\,  \zeta^{\dot\a}\, T\ .\label{tsuper}
\end{equation}

\ni The reducibility relations among the constraints can now be written
in the concise form \cite{us1}
\be
 {\cal P}_\a {\cal T}=0\ . \label{reduceq}
 \ee

\ni In fact, the system has
infinite order reducibility. This can be easily seen from the form
${\cal P}_\a\, {\cal T}=0$ for the  reducibility relations, owing to the
fact that the functions ${\cal P}_\a$ are themselves reducible, since
${\cal P}_\a\, {\cal P}^\a$ gives back the constraints ${\cal T}$. This
infinite order of reducibility implies that a proper BRST treatment
requires an infinite number of ghosts for ghosts
\footnote{Note that, this situation is very nuch similar to the case of
systems with  $\kappa$-symmetry.}.
The construction of the covariant BRST operator is rather cumbersome
problem. Some den progress is made on this problem in \cite{us1}, however, 
thanks
to the fact that the covariant system is critical.

To have an insight into the physical spectrum of the theory, and its
basic interactions, it is sufficient to consider the independent subset
of constraints, at the expense of sacrificing manifest target space
supersymmetry. For example, we can choose the following set of
independent constraints \cite{us1}
\begin{equation}
T = -\ft12 \del X^{\alpha\dot\alpha}\, \del X_{\alpha\dot\alpha} -
p_\alpha\, \del\theta^\alpha\ , \qquad  G^{\dot1} = - p_\alpha\, \del
X^{\alpha\dot1}\ ,\label{n1matcur}
\end{equation}

\ni which in fact generate a subalgebra of the twisted $N=2$
superconformal algebra. Using \eqn{reduceq}, we can write the remaining
constraints, {\it i.e.~}the dependent ones, as linear functions of the
independent constraints
\footnote{Although the massless states can be shown to be annihilated by
the dependent constraints as well, it turns out that there are massive
operators  with standard ghost structure which do not seem to be
annihilated by them \cite{us1}. Establishing the equivalence of the
massive spectra of the reducible and the irreducible systems
would require the analysis of the full cohomology and interactions,
including the physical states with non-standard ghost structure.}.

The BRST operator for the reducible system $(T, G^{\1d})$ can be easily
constructed.  We introduce the anticommuting ghosts $(b,c)$
and the commuting ghosts $(r,s)$ for $T$ and $G^{\dot1}$ respectively. 
The commuting ghosts $(r, s)$ are bosonized, {\it i.e.} $r=\del\xi\,
e^{-\phi}$, $s=\eta\, e^{\phi}$. In terms of these fields, the BRST
operator $Q$ is given by \cite{us1}
\begin{eqnarray}
Q&=& c \Big(-\ft12 \del
X^{\a\dot\a}\, \del X_{\a\dot\a} - p_\a\, \del\theta^\a - b\, \del c
-\ft12 (\del\phi)^2 -\ft32 \del^2\phi - \eta\, \del \xi\Big) \nonumber\\
&+& \eta\, e^{\phi}\, p_\a\,\del X^{\a\dot1}\ .\label{n1brst}
\end{eqnarray}

The theory has spacetime supersymmetry, generated by \cite{us1}
\begin{eqnarray}
q^\a &=& \oint p^\a\ ,\nonumber\\
q^{\dot1} &=&\oint \theta_\a\, \del X^{\a\dot1}\ ,\qquad
q^{\dot2} =\oint \theta_\a\, \del X^{\a\dot2} + b\,\eta\,e^{\phi}\ .
\label{n1susygen}
\end{eqnarray}

\ni The somewhat unusual ghost terms in $q^{\dot2}$ are necessary for the
generator to anti-commute with the BRST operator. It is straightforward to
verify that these supercharges generate the usual $N=1$ spacetime
superalgebra
\begin{equation}
\{q_\alpha,q_\beta\}=0= \{q^{\dot\alpha},q^{\dot\beta} \},\qquad
\{q^\alpha,q^{\dot\alpha} \}= P^{\alpha\dot\alpha} \ ,\label{n1susyalg}
\end{equation}

\ni where $P^{\alpha\dot\alpha}=\oint \del X^{\alpha\dot\alpha}$ is the
spacetime translation operator.

Since the zero mode of $\xi$ is not included in the Hilbert space of
physical states, there exists a BRST non-trivial picture-changing operator
$Z=\{Q, \xi\}$ which can give new BRST non-trivial physical operators when
normal ordered with others.  Explicitly, it takes the form \cite{us1}
\begin{equation}
Z=c\,\del \xi + p_\a\, \del X^{\a\dot1} e^{\phi}\ . \label{n1pic}
\end{equation}

\ni Unlike the picture-changing operator in the usual $N=1$ NSR
superstring, this operator has no inverse.

Let us now consider the physical spectrum with standard ghost structure.
There are two massless operators \cite{us1}
\begin{equation}
V=c\, e^{-\phi}\, e^{ip\cdot X}\ ,\qquad
\Psi = h_\a\, c\, e^{-\phi}\, \theta^\a\, e^{ip\cdot X}\ ,\label{n1mass0}
\end{equation}

\ni which are physical provided with mass-shell condition $p^{\a\dot\a}\,
p_{\a\dot\a} = 0$ and spinor polarisation condition $p^{\a\dot1}\, h_a =0$.
The non-triviality of these operators can be established by the fact that
the conjugates of these operators with respect to the following
non-vanishing inner product
\begin{equation}
\Big\langle \del^2c\,\del c\, c\, e^{-3\phi}\, \theta^2 \Big\rangle
\label{n1innpro}
\end{equation}

\ni are also annihilated by the BRST operator. The bosonic operator $V$
and the fermionic operator $\Psi$ form a supermultiplet under the $N=1$
spacetime supersymmetric transformation.  The associated spacetime
fields $\phi$ and $\psi_\a$ transform as
\begin{equation}
\delta\phi = \epsilon_\a\, \psi^\a\, \qquad
\delta\psi_\a = \epsilon^{\dot\a}\, \del_{\a\dot\a}\, \phi\ .
\end{equation}

We can build only one three-point amplitude among the massless
operators, namely \cite{us1}
\begin{equation}
\Big\langle V(z_1)\,\, \Psi(z_2)\,\, \Psi(z_3)\Big\rangle =c_{23}\ ,
\label{n13pf}
\end{equation}

\ni where $b_{ij}$ is defined by
\begin{equation}
b_{ij} =  h_{(i)\a}\,h^\a_{(j)}\ .\label{n13pf1}
\end{equation}

\ni From this, we can deduce that the $V$ operator describes a spacetime
scalar whilst the $\Psi$ operator describes a spacetime chiral
spin-$\ft12$ fermion.  Note that this is quite different from the case
of the $N=2$ string where there is only a massless boson and although it
is ostensibly a scalar, it is in fact, as emerges from the study of the
three-point amplitudes, a prepotential for self-dual Yang-Mills or
gravity.

With the one insertion of the picture-changing operator, we
can build a four-point function which vanishes for kinematic reasons
\cite{us1}:
\begin{equation}
\Big\langle ZV\, \Psi\, \oint b\Psi\, \Psi \Big\rangle  =
(u\, b_{12}\, b_{34} + s\, b_{13}\, b_{24}){\Gamma(-\ft12 s)
\, \Gamma(-\ft12 t)\over \Gamma(\ft12 u)}\ ,\label{4pfmass0}
\end{equation}

\ni where $s$, $t$, and $u$ are the Mandelstam variables and $h_{(1)}^\a
= p^{\a\dot1}_{(1)}$.  The vanishing of the kinematic term, {\it i.e.}
$u\, b_{12}\, b_{34} + s\, b_{13}\, b_{24}=0$, is a straightforward
consequence of the mass-shell condition of the \ni operators and
momentum conservation of the four-point amplitude \cite{lp}. It might
seem that the vanishing of the this four-point amplitude should be
automatically implied by the statistics of the operators since there is
an odd number of fermions. However, as shown in \cite{us1}, the
picture-changing operator has spacetime fermionic statistics.   In fact,
that the four-point amplitude \eqn{4pfmass0} vanishes only on-shell, for
kinematic reasons,  already implies that the picture changer $Z$ is a
fermion. Thus the picture changing of a physical operator changes its
spacetime statistics and hence does not establish the equivalence
between the two. On the other hand, since $Z^2=(ZZ)$ becomes a spacetime
bosonic operator,  we can use $Z^2$ to identify the physical states with
different pictures.

Thus, we have a total of four massless operators,
namely $V$, $ZV$ and their supersymmetric partners.  $V$ and its
superpartner $\Psi$ have standard ghost structure; $ZV$ and its
superpartner $Z\Psi$ have non-standard ghost structures. 

So far we have discussed the massless physical operators. There are also
infinitely many massive states. The tachyonic type massive operators,
{\it i.e.~} those that become pure exponentials after bosonizing the
fermionic fields, are relatively easy to obtain, and they have been
discussed at length in \cite{us1}. An example of such massive operators
is as follows
\be
V_n= c(\del^np)^2\cdots p^2\, e^{n\phi}\, e^{ip\cdot X}\ ,\qquad
{\cal M}^2 =(n+1)(n+2)\ , \label{n1masstats}
\ee

\ni where $p^2 = p_\a\, p^\a$. These operators correspond to physical
states, provided the mass-shell condition is satisfied. Furthermore, they
all have non-standard ghost structures. From these operators, we can
build non-vanishing four-point amplitudes, which implies the existence
of further massive operators in the physical spectrum.

In summary, we  emphasize that the model has $n=1$ supersymmetry in the
critical $2+2$ dimensional spacetime. It describes two massless scalar
supermultiplets, in addition to an infinite tower of massive states.
Examining their interactions, however, we find that they do not
correspond to those of self-dual supergravity.
\vspace{0.8cm}
\subsection{The New $n=0$ Model}

This model is described by the following set of currents \cite{us2}
\begin{eqnarray}
T&=& -\ft12 \del X^{\alpha\dot\alpha}\del X_{\alpha\dot\alpha}
     -p_\alpha \del \theta^\alpha \ , \qquad J =\del (\theta_\alpha
      \theta^\alpha) \ , \nonumber\\
       G^{\dot 1} &=& p_{\alpha}\del X^{\alpha\dot 1} \ , \qquad
       {\widetilde G}^{\dot 1}=\theta_\alpha \del X^{\alpha\dot 1}\ .
\label{newalg}
\end{eqnarray}

\ni It is easy to see that the currents $(T, G^{\dot 1}, {\widetilde
G}^{\dot 1}, J)$ have spins $(2,2,1,1)$. In addition to the standard
OPEs of $T$ with $(T,J,G^{\dot 1},{\widetilde G}^{\dot 1})$, the only
non-vanishing OPE is
\begin{equation}
J(z)\, G^{\dot 1} (w) \sim {2{\widetilde G}^{\dot 1}\over (z-w)^2}+
         {\del {\widetilde G}^{\dot 1}\over (z-w)} \ .
\end{equation}

\ni This algebra is related to the small $N=4$ superconformal algebra,
not directly as a subalgebra, but in the following way. The subset of
currents $T, G^{\dot 1}, \widetilde G^{\dot 1}$ and $J_-$ in
(\ref{n4alg}) form a critical closed algebra. However these currents
form a reducible set. To achieve irreducibility, we simply differentiate
the current $J_-$, thereby obtaining the set of currents given in
(\ref{newalg}). Note that taking the derivative of $J_-$ still gives a
primary current with the same anomaly contribution, since $12s^2-12s+2$
takes the same value for $s=0$ and $s=1$.

  To proceed with the BRST quantisation of the model, we introduce the
fermionic ghost fields $(c,b)$ and $(\gamma,\beta)$ for the currents $T$
and $J$, and the bosonic ghost fields $(s,r)$ and $({\tilde s},{\tilde
r})$ for $G^{\dot 1}$ and ${\widetilde G}^{\dot 1}$. It is necessary to
bosonize the commuting ghosts, by writing $s=\eta e^\phi$, $r=\del \xi
e^{-\phi}$, $\tilde s=\tilde \eta e^{\tilde \phi}$ and $\tilde r=\del
\tilde\xi e^{-\tilde\phi}$. The BRST operator for the model is then
given by \cite{us2}
\begin{eqnarray}
 Q&=& \oint c\Big( -\ft12 \del X_{\alpha\dot\alpha}\del
X^{\alpha\dot\alpha} -p_\alpha \del \theta^\alpha-\ft12 ( \del \phi)^2
-\ft12 (\del\tilde \phi)^2 -\ft32 \del^2 \phi -\ft12 \del^2 \tilde\phi
\nonumber \\ &&  -\eta\del\xi -\tilde\eta \del \tilde \xi -b\del c
-\beta\del\gamma\Big)
\label{brst} \\
&&+\eta e^\phi p_\alpha \del
X^{\alpha \dot 1} + \tilde \eta e^{\tilde\phi} \theta_\alpha \del
X^{\alpha \dot 1} + \del \gamma \Big( \ft12 \theta^\alpha\theta_\alpha
-\del\tilde\xi\eta e^{\phi-\tilde\phi}\Big)\ .  \nonumber
\end{eqnarray}

\ni Since the zero modes of $\xi$ and $\tilde\xi$ do not appear in the
BRST operator, there exist BRST non-trivial picture-changing operators
\cite{us2}:
\begin{eqnarray}
Z_\xi &=& \{ Q, \xi\}=c\del\xi+e^\phi p_\alpha\del X^{\alpha\dot 1}
-\del \gamma \del\tilde\xi e^{\phi-\tilde\phi}\ , \nonumber\\
Z_{\tilde\xi} &=& \{ Q, \tilde\xi\}=c\del\tilde\xi + e^{\tilde\phi}
\theta_\alpha\del X^{\alpha\dot 1}\ . \label{pic}
\end{eqnarray}

\ni It turns out that these two picture changers are not invertible.
Thus, one has the option of including the zero modes of $\xi$ and
$\tilde\xi$ in the Hilbert space of physical states. This would not be
true for a case where the picture changers were invertible. Under these
circumstances, the inclusion of the zero modes would mean that all
physical states would become trivial, since $|{\rm phys}\rangle =Q(\xi
Z_\xi^{-1} |{\rm phys}\rangle)$. In \cite{us2}, we chose to exclude the
zero modes of $\xi$ and $\tilde\xi$ from the Hilbert space. It is
interesting to note that in this model the zero mode of the ghost field
$\gamma$ for the spin--1 current is also absent in the BRST operator. If
one excludes this zero mode from the Hilbert space, one can then
introduce the corresponding picture-changing operator $Z_\gamma =
\{Q,\gamma\}=c\del \gamma$. In \cite{us2}, we  indeed chose to exclude
the zero mode of $\gamma$.

In order to discuss the cohomology of the BRST operator \eqn{brst}, it
is convenient first to define an inner product in the Hilbert space.
Since the zero modes of the $\xi,\tilde\xi$ and $\gamma$ are excluded,
the inner product is given by
\be
\langle \del^2 c\,\del c\, c\, \theta^\alpha\theta_\alpha\,
e^{-3\phi-\tilde\phi}\rangle=1\ . \label{innerp}
\ee

Let us first discuss the spectrum of massless states in the Neveu-Schwarz
sector. The simplest such state is given by \cite{us2}
\be
 V=c\, e^{-\phi-\tilde\phi} e^{ip\cdot X} \ . \label{massless}
\ee

\ni As in the case of the $N=2$ string discussed in \cite{lp}, since the
picture-changing operators are not invertible the massless states in
different pictures cannot necessarily all be identified. In fact, the
picture changers annihilate the massless operators such as
(\ref{massless}) when the momentum $p^{\alpha\dot 1}$ is zero. However,
massless operators in other pictures still exist at momentum
$p^{\alpha\dot 1}=0$. For example, in the same picture as the physical
operator $Z_{\tilde \xi} V$ that vanishes at $p^{\alpha\dot 1}=0$ is a
physical operator that is non-vanishing for all on-shell momenta, namely
\cite{us2}
\be
\Psi=h_\alpha\, c\,\theta^\alpha\, e^{-\phi}\, e^{ip\cdot X}\ ,
\label{psi}
\ee

\ni which is physical provided that $p^{\alpha\dot 1}\, h_\alpha=0$ and
$p_{\alpha\dot\alpha} p^{\alpha\dot\alpha}=0$. In fact, $Z_{\tilde \xi}
V$ is nothing but $\Psi$ with the polarisation condition solved by
writing $h^\alpha=p^{\alpha\dot 1}$. However, we can choose instead to
solve the polarisation condition by writing $h^\alpha=p^{\alpha\dot 2}$,
which is non-vanishing even when $p^{\alpha\dot 1}=0$. Thus, the
operators $V$ and $\Psi$ cannot be identified under picture changing
when $p^{\alpha\dot1}=0$. In fact when $p^{\alpha\dot1}=0$ there is
another independent solution for $\Psi$, since the polarisation
condition becomes empty in this case.  A convenient way to describe the
physical states is in terms of $\Psi$ given in \eqn{psi}, with the
polarisation condition re-written in the covariant form
$p^{\alpha\dot\alpha}\, h_\alpha=0$, together with a further physical
operator which is defined only when $p^{\alpha\dot1}=0$.  In this
description, the physical operator $\Psi$ is defined for all on-shell
momenta.

     If one adopts the traditional viewpoint that physical operators
related by picture changers describe the same physical degree of
freedom, one would then interpret the spectrum as containing a massless
operator \eqn{massless}, together with an infinite number of massless
operators that are subject to the further constraint $p^{\alpha\dot1}=0$
on the on-shell momentum
\footnote{It should be emphasized that the possibility of having
$p^{\alpha\dot 1}=0$ while $p^{\alpha\dot 2}\ne 0$ is a consequence of
our having chosen a real structure on the $(2,2)$ spacetime
\cite{us1,lp}, rather than the more customary complex structure
\cite{ov}.}.
This viewpoint is not altogether satisfactory in a
case such as ours, where the picture changing operators are not
invertible. An alternative, and moreover covariant, viewpoint is that
the physical operators in different pictures, such as $V$ and $\Psi$,
should be viewed as independent. At first sight one might think that
this description leads to an infinite number of massless operators.
However, as shown in \cite{us2}, the interactions of all the physical
operators can be effectively described by the interaction of just the
two operators $V$ and $\Psi$.

Thus the theory effectively reduces to one with just two massless
operators, one a scalar and the other a spinorial bosonic operator.

As for the massive states, an infinite tower of them exist, and they
have been discussed in \cite{us2}. They all have postive mass, and
non-standard ghost structure
\footnote{By considering the interactions, one can deduce the existence of
an infinite tower of massive states with standard ghost-structure as
well \cite{us2}.}.
A typical such tower is given by \cite{us2}
\be
V_n = c\, (\del^n p)^2\cdots p^2\,
           e^{n\phi-(n+2)\tilde\phi}\,
           \del^{2n+2}\gamma\cdots \del\gamma\, e^{ip\cdot X}\ ,
\label{massive}
\ee

\ni where $n> -1$ and the mass is given by ${\cal M}^2=(2n+2)(2n+3)$.
 For subtleties concerning the exclusion of the zero-mode of
the $\g$ field in the Hilbert space of physical states, and the nature
of the picture-changin operators in massless versus massive sector of
the theory, we refer the reader to \cite{us2}.

As for the interactions, there is one three-point interaction between
the massless operators, namely \cite{us2}
\be
\Big \langle \Psi(z_1, p_{(1)})\, \Psi(z_2, p_{(2)})\, V(z_3, p_{(3)})
\Big\rangle = h_{(1)\alpha}\, h_{(2)}^\alpha \ .\label{3pf2}
\ee

\ni Note that this three-point amplitude is manifestly Lorentz invariant.
There are also an infinite number of massless physical operators with
different pictures in the spectrum, and they can all be expressed in a
covariant way.  As one steps through the picture numbers, the character
of the physical operators alternates between scalar and spinorial.  The
three-point interactions of all these operators lead only to the one
amplitude given by \eqn{3pf2}.  In view of their equivalent
interactions, all the scalar operators can be identified and all the
spinorial operators can be identified.

 The massless spectrum can thus be effectively described by the scalar
operator \eqn{massless} and the spinorial operator \eqn{psi}.  All
four-point and higher amplitudes vanish.

Although the theory contains an infinite tower of physical operators,
the massless sector and its interactions are remarkably simple.   In
particular, although all the massive physical operators break Lorentz
invariance, the massless operators and their interactions have manifest
spacetime Lorentz invariance.  If we associate spacetime fields $\phi$
and $\psi_\alpha$ with the physical operators $V$ and $\Psi$, it follows
from the three-point amplitude \eqn{3pf2} that we can write the field
equations \cite{us2}:
\be
\del_{\alpha\dot\alpha}\del^{\alpha\dot\alpha}\phi=\psi^\alpha\,
\psi_\alpha\, \qquad \del_{\alpha\dot\alpha}\psi^\alpha =
\psi^\alpha\del_{\alpha\dot \alpha}\phi\ .\label{feom}
\ee

We have suppressed Chan-Paton group theory factors that must be
introduced in order for the three-point amplitude to be non-vanishing in
the open string. It is easy to see even from the kinetic terms in the
field equations \eqn{feom} that there is no associated Lagrangian.  Note
that there is no undifferentiated $\phi$ field, owing to the fact that
the theory is invariant under the transformation $\phi\longrightarrow
\phi + {\rm const.}$ It is of interest to obtain the higher-point
amplitudes from the field equations \eqn{feom}, which should be zero
if they are to reproduce the string interactions.

In summary, the new $n=0$ model has a massless scalar and fermion, in
addition to an infinite tower of massive particles. However, the model
lacks spacetime supersymmetry. Moreover, while the massless fields have
interesting interactions, for which we can write down the field
equations not derivable from a Lagrangian, the model does not seem to
describe the interactions of self-dual gravity.
\vspace{8mm}

\noindent{\bf Acknowledgements}

\vspace{4mm}

The work described in this paper owes greatly to H. L\"u and C.N. Pope,
who are gratefully acknowledged. This work was supported in part by the
National Science Foundation, under grant PHY--9411543.

\vfill\eject

\end{document}